# Decomposition process in a FeAuPd alloy nanostructured by severe plastic deformation

X. Sauvage*[1], A. Chbihi[1], D. Gunderov[2], E.V. Belozerov[3], A.G. Popov[3]

1- University of Rouen, CNRS UMR 6634, Groupe de Physique des Matériaux, Faculté des Sciences, BP12, 76801 Saint-Etienne du Rouvray, France

2- Institute for Physics of Advanced Materials, Ufa State Aviation Technical University, K. Marx 12, Ufa 450000, Russia

3- Institute of Metal Physics, 18 Kovalevskaya str., Ekaterinburg, 620219, Russia

*corresponding author: Xavier Sauvage
xavier.sauvage@univ-rouen.fr
Tel : + 33 2 32 95 51 42          Fax : + 33 2 32 95 50 32

**Abstract**
The decomposition process mechanisms have been investigated in a Fe50Au25Pd25 (at.%) alloy processed by severe plastic deformation. Phases were characterized by X-ray diffraction and microstructures were observed using transmission electron microscopy. In the coarse grain alloy homogenized and aged at 450°C, the bcc $\alpha$-Fe and fcc AuPd phases nucleate in the fcc supersaturated solid solution and grow by a discontinuous precipitation process resulting in a typical lamellar structure. The grain size of the homogenized FeAuPd alloy was reduced in a range of 50 to 100nm by high pressure torsion. Aging at 450°C this nanostructure leads to the decomposition of the solid solution into an equi-axed microstructure. The grain growth is very limited during aging and the grain size remains under 100nm. The combination of two phases with different crystallographic structures (bcc $\alpha$-Fe and fcc AuPd ) and of the nanoscaled grain size gives rise to a significant hardening of the alloy.

**1. Introduction**

It is now well demonstrated that Severe Plastic Deformation (SPD) processes are able to produce fully dense bulk nanostructured materials in large quantity. SPD related techniques like Equal Channel Angular Pressing (ECAP) [1], High Pressure Torsion (HPT) [1] or Accumulative Roll Bounding [2] are usually applied to metals and alloys to produce ultrafine grained structures. The grain size is typically in a range of 100 to 300 nm for commercially pure metals [1] and could be less than 20 nm for some specific alloys [3, 4, 5, 6, 7]. It is well known that the smaller the grain size, the higher the yield stress (Hall and Petch law [8, 9]), thus SPD processed materials usually exhibit a very high strength comparing to their coarse grained counterparts. On the other hand, most of them exhibit a low level of ductility [10, 11], with a uniform elongation to failure typically less than 5%. This lack of ductility is a critical issue and up to now it has strongly limited the application of bulk nanostructured materials produced by SPD as structural components. However numerous research teams around the world have focused their activities on this issue and there is now some great hopes to achieve nanostructures with an excellent combination of high strength and good ductility. Thus, E. Ma and co-authors have recently suggested and also demonstrated that nanoscaled precipitates together with ultrafine grains is a suitable route to reach this goal [12, 13]. The physical reasons could be summarized as follow: in single phase ultrafine grains, dislocations are pinned by grain boundaries and there is no strain hardening because of the lack of dislocation sources. But if there are some nanoscaled precipitates within the grains, they could pin gliding dislocations and contribute to the formation of dislocation sources which gives rise to some potential strain hardening and ductility. However, controlling the precipitation of second phase particles in an alloy nanostructured by SPD could be a real challenge. Indeed, during the aging treatment necessary for the nucleation of precipitates, recovery, recrystallization and grain growth may occur. Moreover, there are so many defects in such nanostructured materials that heterogeneous precipitation along grain boundaries or dislocations is more likely to occur than homogeneous precipitation. The aim of this paper is to bring further knowledge on phase transformation mechanisms in SPD processed metallic alloys. These features have been investigated in a FeAuPd alloy processed by HPT. The precipitation of the α-Fe bcc phase from the fcc supersaturated solid solution was investigated by X-ray diffraction and Transmission Electron Microscopy (TEM). The microstructure of the aged alloy was compared to that of the HPT alloy aged in similar conditions.

## 2. Experimental

The alloy investigated in present study is a Fe50-Au25-Pd25 (at.%) that was cast from high purity Fe (99.5%) and a Au50Pd50 (at.%) alloy. The as-cast material was homogenized at 900°C during 6h in Ar atmosphere and subsequently quenched in ice brine. A part of the homogenized alloy was aged during 30h at 450°C and the other part was used for SPD experiments. The material was processed by HPT at room temperature under a pressure of 6 GPa. HPT samples (10 mm in diameter discs with a thickness of 0.3 mm) were subjected to 5 revolutions (0.2 rev/min) and then aged during 30h at 450°C.

Crystallographic phases were characterized by X-ray diffraction (XRD). Spectra were recorded with a Brucker D8 system in Bragg-Brentano θ-2θ geometry. The X-ray generator was equipped with a Co anticathode, using Co (Kα) radiation (λ = 0.17909 nm). The measurements were performed so that the out-of-plane component was the torsion axis of the sample processed by HPT.

Microstructures were characterized by TEM. Observations were performed with a JEOL 2000FX microscope operating at 200 kV. TEM samples were prepared by ion milling (PIPS-GATAN 691, 3kV, beam angle 4°, 300K). Samples with a diameter of 3mm were cut in the discs processed by HPT at a distance of 3 mm from the disc center (corresponding to a shear strain of about γ ≈ 300). Thus, TEM investigations were performed with the electron beam parallel to the torsion axis.

The Vickers microhardness of the alloy was measured with a BUEHLER Micromet 2003 machine, using a 300g load (2.94 N).

## 3. Results and discussions
### *Decomposition process in the coarse-grained FeAuPd alloy*

The XRD spectrum of the homogenized FeAuPd alloy is displayed in the Fig.1(a). It clearly shows that the probed sample contains only one fcc disordered phase with all elements in solid solution. The lattice parameter of this phase calculated from these data is : 0.510 nm (± 0.001). After aging at 450°C during 30h (Fig.1(b)), an additional phase is detected indicating that the solid solution decomposed. The new peak corresponds to the (110) diffraction peak of the bcc α-Fe phase (lattice parameter: 0.287 nm ± 0.001). The TEM bright field image (Fig.1(c)) clearly shows the two-phase microstructure. Obviously some discontinuous precipitation occurred resulting in a typical lamellar structure [14]. In agreement with XRD data, the bright lamellae have a bcc crystallographic structure, and using X-ray Energy Dispersive Spectroscopy (EDS), only Fe was detected (data not shown here). These bcc lamellae are embedded in the disordered fcc AuPd matrix (Fe was not detected in this phase by EDS). The lamellae thickness is in a range of 10 to 30 nm, while their length is in a range of 50 to 300 nm. The lamellae have a wavy shape and it is therefore difficult to fully image one of them orientated perpendicular to the electron beam. Anyway, some attempts were carried out and a lamella imaged with such an orientation is shown in the Fig.2. From such images, the average width of lamellae (third dimension) was estimated in a range of 50 to 100 nm. It is interesting to note that the nucleation and growth of these lamellae induce a significant hardening of the alloy: from 2 GPa after the homogenization treatment, up to 3.7 GPa after aging (see table 1).

*Table 1*
*Microhardness of the FeAuPd alloy after homogenization, homogenization + aging at 450°C during 30h, homogenization + 5 revolutions by HPT, homogenization + 5 revolutions by HPT + aging at 450°C during 30h.*

|  | **Homogenized** | **Aged** | **HPT** | **HPT+aged** |
|---|---|---|---|---|
| HV (GPa) | 2 ±0.1 | 3.7 ±0.1 | 3.4 ±0.1 | 4.7 ±0.1 |

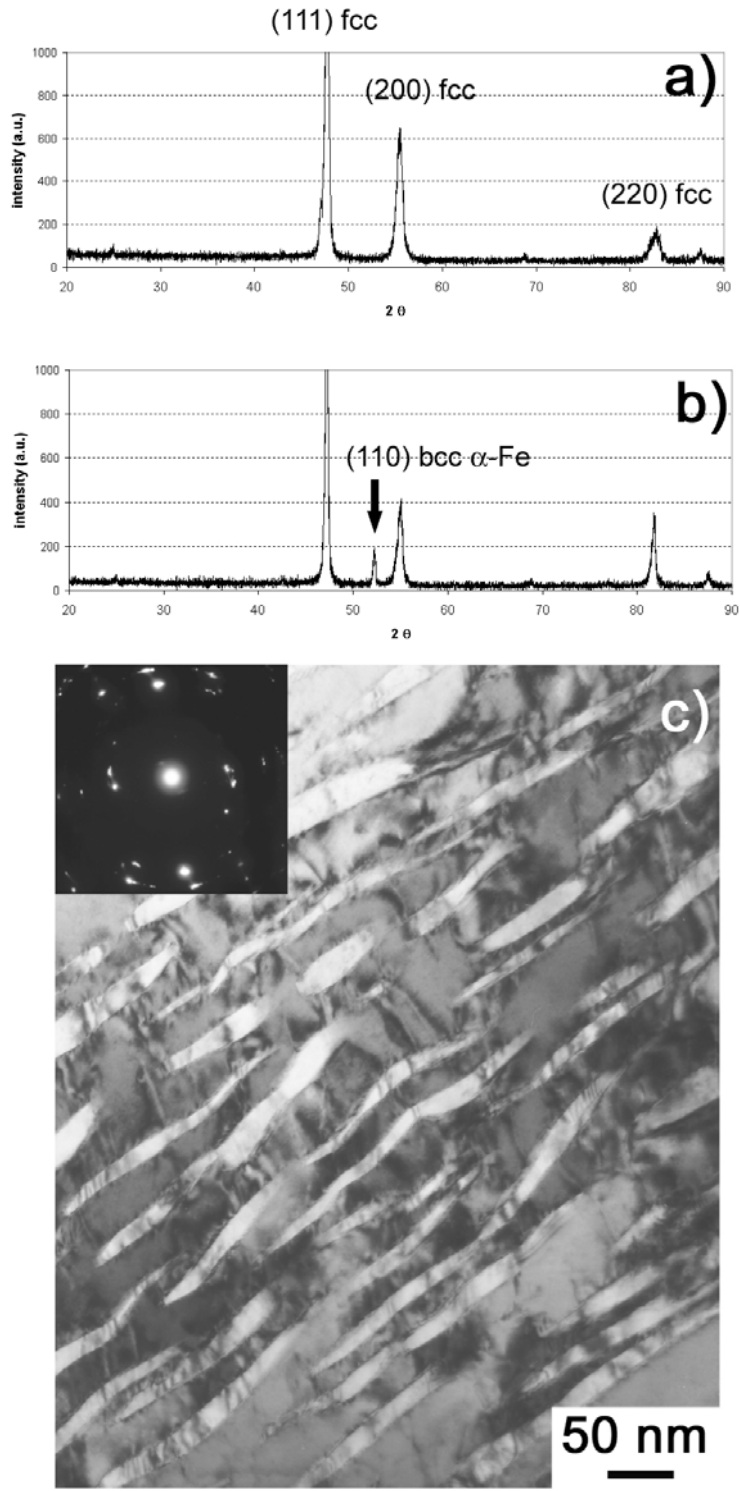

*Figure 1*
*(a) X-ray diffraction spectrum of the homogenized FeAuPd alloy showing that there is a single fcc phase. (b) X-ray diffraction of the non-deformed FeAuPd alloy after aging at 450°C during 30h showing that the a-Fe bcc has nucleated. (c) TEM bright field image and corresponding SAED pattern (aperture size 2μm) showing the microstructure of non-deformed FeAuPd alloy after aging at 450°C during 30h.*

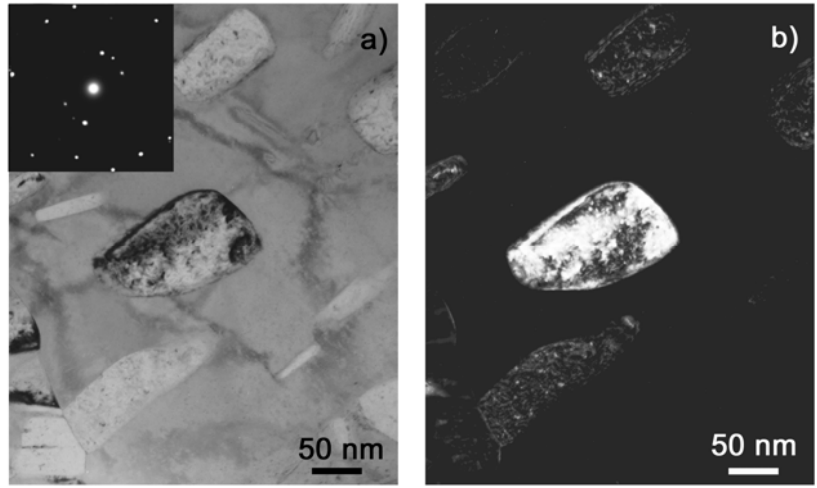

*Figure 2*
*(a) High magnification TEM bright field image and corresponding SAED pattern (aperture size 2µm) showing the cross section of a rod shaped bcc α-Fe precipitate in the non-deformed FeAuPd alloy after aging at 450°C during 30h. (b) Corresponding dark field image were only the α-Fe precipitate is imaged.*

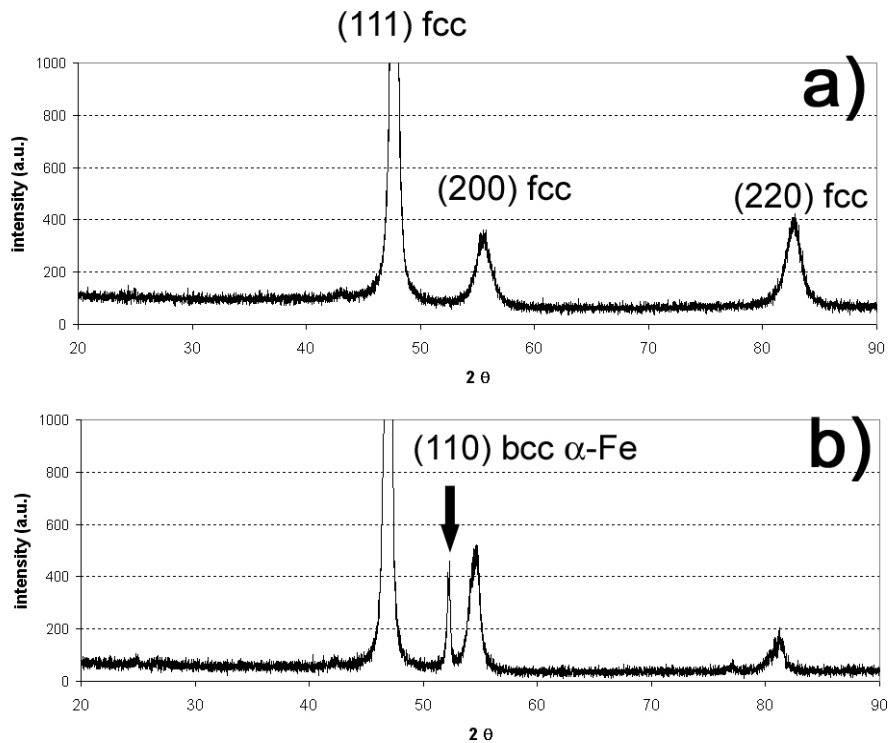

*Figure 3*
*(a) X-ray diffraction spectrum of the homogenized FeAuPd alloy after HPT processing showing that there is a single fcc phase. (b) X-ray diffraction of the homogenized FeAuPd alloy after HPT processing followed by subsequent aging at 450°C during 30h showing that the α-Fe bcc has nucleated.*

## Microstructure of the FeAuPd alloy processed by HPT

The XRD spectrum of the homogenized FeAuPd alloy processed by HPT is displayed in the Fig.3(a). Comparing to the non-deformed alloy (Fig. 1(a)), there is a significant broadening of the diffraction peaks. This feature is commonly observed in SPD metals and results from grain size reduction and internal stresses [1]. As shown on the TEM bright field image, the grain size is indeed very small after HPT (Fig. 4). The selected area electron diffraction pattern recorded with an aperture of only 2μm exhibits Debye-Scherrer rings typical of nanocrystalline structures with high angle grain boundaries. The dark field image (Fig. 4(b)) obtained by selecting a part of the more intense ring clearly shows that the grain size is in a range of 50 to 100nm. As a result of this nanostructuration, the hardness has significantly increased up to 3.4 GPa, thus reaching a value close to that obtained after aging and precipitation in the non-deformed material. Even if the level of deformation is much higher in the outer part of the HPT disc [1], it is worth noticing that the hardness does not fluctuate significantly along the diameter of the sample.

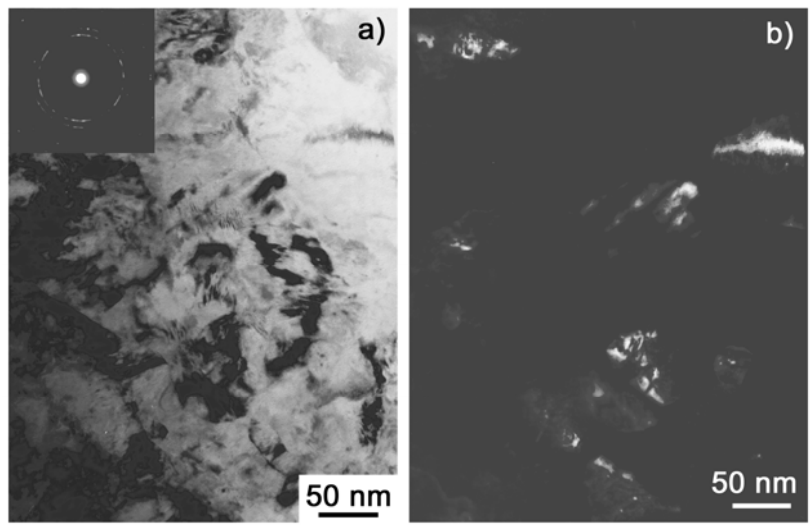

*Figure 4*
*(a) TEM bright field image and corresponding SAED pattern (aperture size 2μm) of the microstructure of the homogenized FeAuPd alloy processed by HPT. (b) Corresponding dark field image showing few isolated fcc grains.*

## Decomposition process in the nanocrystalline FeAuPd alloy

The FeAuPd alloy processed by HPT was aged at 450°C during 30h and the XRD spectrum (Fig 3(b)) does not show any significant difference comparing to the non-deformed alloy (Fig 1(b)). Similarly, decomposition occurred during the aging treatment and the fcc matrix is detected together with the bcc α-Fe phase. However, as exhibited on the bright field TEM picture (Fig. 5(a)), the microstructure is very different. The grain size did not change significantly during aging (in a range of 70 to 100 nm) and obviously bcc α-Fe grains that have nucleated and grew are not lamella shaped. The high magnification TEM bright field image of the Fig. 5(b) displays a large bcc α-Fe grain (brightly imaged). This large grain was selected because smaller grains were difficult to image. Anyway, it shows that bcc α-Fe grains are nanoscaled and almost equiaxed. Thus, it seems that during the aging treatment heterogeneous nucleation occurred along grain boundaries of the nanocrystalline fcc solid solution produced by SPD. It is interesting to note that the precipitation of this second phase induces a significant increase of the hardness up to 4.7 GPa, which is 1 GPa higher than the hardness of the non-deformed material after similar aging treatment. As noticed for the HPT sample, the hardness was fairly constant across the diameter of the sample.

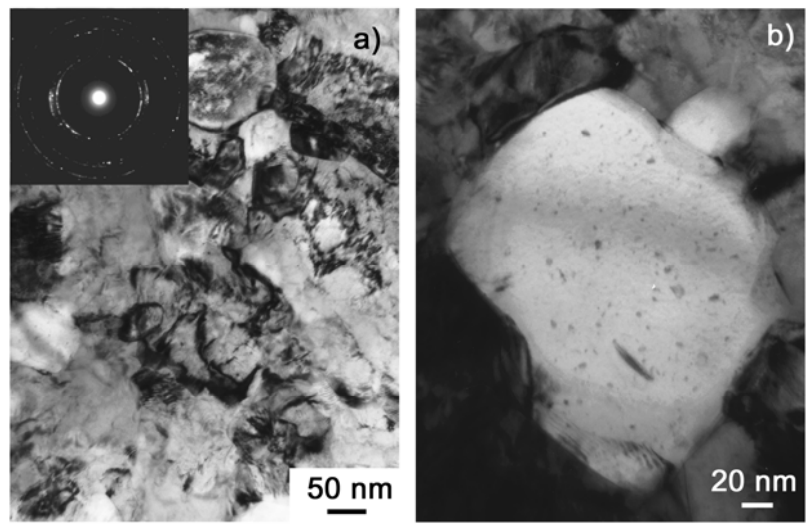

*Figure 5*
*(a) TEM bright field image and corresponding SAED pattern (aperture 3µm) of the microstructure of the homogenized FeAuPd alloy processed by HPT and subsequently aged at 450°C during 30h. (b) High magnification TEM bright field image of an equi-axed bcc α-Fe grain that has precipitated within the nanostructure.*

## 4. Discussion

The decomposition process in the coarse grained FeAuPd alloy is clearly a discontinuous precipitation regime resulting in a typical lamellar microstructure thanks to the redistribution of Au, Pd and Fe atoms (Fig. 6(a)). The high volume fraction of lamella shaped bcc α-Fe precipitates provide some efficient obstacles for dislocation glide and this leads to a significant increase of the hardness (table 1). In the nanostructured alloy, the situation is very different because there are numerous high angle grain boundaries (Fig. 4) that could act as nucleation sites. Moreover, as schematically represented on the Fig.6 (b), the grain size is close to the typical length scale of the lamellar structure that grows in the coarse grain alloy. Thus theoretically, if a bcc α-Fe lamella grows from one grain boundary, it would simply split a grain. It seems however that this situation never occurs since only equi-axed grains were observed by TEM (Fig. 5). This feature could be attributed to the higher atomic mobility of solute elements along grain boundaries: once a grain nucleates on a grain boundary, it is simply fed by atoms migrating along this grain boundary (Fig. 6(c)). This scenario is completely different from the decomposition process that occurs in the coarse grain material, and is the consequence of the nanoscaled structure that was achieved by SPD. It is also interesting to note that there is no significant grain growth during the aging treatment (Fig. 4 and Fig. 5). This indicates that even if there is a significant atomic mobility (otherwise precipitation would not occurred), a small grain can be kept upon aging and precipitation. In the present case, there is a competition between decomposition (bcc α-Fe and fcc AuPd grain growth) and coarsening to reduce the interfacial energy. Since the microstructure is a mixture of two phases with nanoscaled grains, grain boundaries pin each others and cannot migrate easily. Therefore after ageing, even if the microstructure has completely changed (full redistribution of Fe atoms), the small grain size was kept. Finally, this combination of nanoscaled grains together with a mixture of two phases with different crystallographic structures gives rise to a much higher hardness than the non-deformed alloy (table 1).

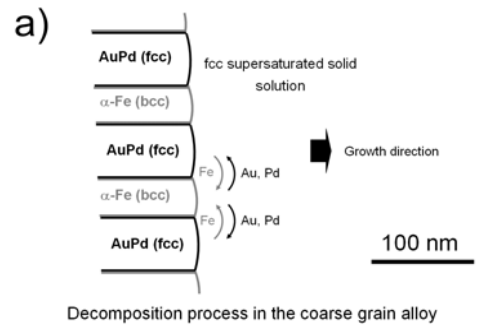
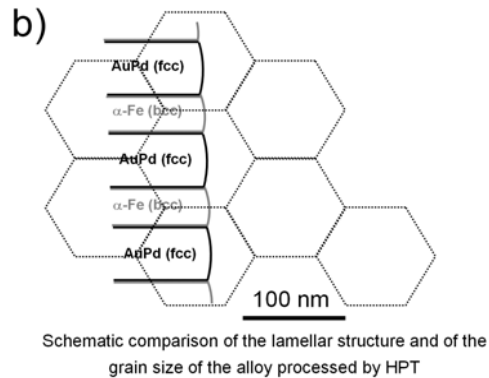
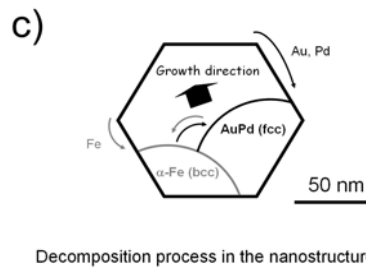

*Figure 6*
*(a) Schematic representation of the decomposition process in the coarse grained alloy resulting in a lamellar structure; (b) Comparison of the grain size after HPT with the typical length scale of the lamellar structure; (c) Schematic representation of the decomposition process in the nanostructured alloy resulting in an equi-axed structure.*

## 5. Conclusions

i) In the coarse grain Fe50Au25Pd25 alloy homogenized at 900°C, the fcc solid solution is transformed into a two phase microstructure during aging at 450°C (fcc AuPd and bcc α-Fe). This is a lamellar structure resulting from a discontinuous precipitation mechanism.

ii) The grain size of the Fe50Au25Pd25 alloy homogenized at 900°C was reduced in a range of 50 to 100nm by HPT. In this nanostructured alloy, the decomposition process is different and does not result in a lamellar structure.

iii) The alloy processed by HPT and subsequently aged at 450°C is completely phase separated but grains do not significantly grow. The combination of two phases with different crystallographic structures and of the nanoscaled grain size gives rise to a strong hardening.

iv) Thus, this work demonstrates that it is possible to combine precipitation hardening together with nanoscaled grains.